\documentclass[aps,pre,twocolumn,floatfix,superscriptaddress,showpacs,showkeys]{revtex4-1}
\usepackage{epsf,amsmath,amssymb,verbatim,color,multirow,pifont,graphicx,tabularx,cancel,soul,float}

\newcommand{\ro}{\rho_0}
\newcommand{\ps}{P_{S}}
\newcommand{\pw}{P_{W}}
\newcommand{\pr}{P_{R}}

\newcommand{\nubar}{\bar \nu}
\newcommand{\self}{self-consistency }

\begin{document}
\setstcolor{red}
\newcommand{\xnew}[1]{\textcolor{cyan}{#1}}

\title{Critical behavior of a two-step contagion model with multiple seeds}
\author{Wonjun Choi}
\affiliation{CCSS, CTP, and Department of Physics and Astronomy, Seoul National University, Seoul 08826, Korea}
\author{Deokjae Lee}
\affiliation{CCSS, CTP, and Department of Physics and Astronomy, Seoul National University, Seoul 08826, Korea}
\author{B. Kahng}
\email{bkahng@snu.ac.kr}
\affiliation{CCSS, CTP, and Department of Physics and Astronomy, Seoul National University, Seoul 08826, Korea}
\date{\today}

\begin{abstract}
A two-step contagion model with a single seed serves as a cornerstone for understanding the critical behaviors and underlying mechanism of discontinuous percolation transitions induced by cascade dynamics. When the contagion spreads from a single seed, a cluster of infected and recovered nodes grows without any cluster merging process. However, when the contagion starts from multiple seeds of $O(N)$ where $N$ is the system size, a node weakened by a seed can be infected more easily when it is in contact with another node infected by a different pathogen seed. This contagion process can be viewed as a cluster merging process in a percolation model. Here, we show analytically and numerically that when the density of infectious seeds is relatively small but $O(1)$, the epidemic transition is hybrid, exhibiting both continuous and discontinuous behavior, whereas when it is sufficiently large and reaches a critical point, the transition becomes continuous. We determine the full set of critical exponents describing the hybrid and the continuous transitions. Their critical behaviors differ from those in the single-seed case.   
\end{abstract}


\maketitle
\section{Introduction}

Nonequilibrium dynamic transitions driven by cascade dynamics on complex networks have attracted considerable attention recently~\cite{review1,review2, review3}. The spreading of epidemic disease on complex networks~\cite{review_epidemics,watts,dodds,krapivsky, grassberger_pre,althouse_prl, chen, porter,grassberger_nphy, althouse_pnas, janssen, janssen_spinodal,chung, hasegawa1,choi_2016} is an instance, in which a pathogen is transmitted from an infected node (e.g., a person) to a susceptible neighbor, who then becomes infected with a certain probability. If the transmission probability is sufficiently large (small), the pathogen spreads out to a macroscopic scale (remains local). An epidemic transition occurs between these two limits. The extent of spreading also depends on the structure of an underlying network~\cite{review1,barabasi}. When degree distribution of a network is highly heterogeneous, diseases can spread out massively even for a small transmission probability, so that an epidemic transition point  can be zero~\cite{sf}. Information spreading in social media from one page to others may be modeled in a similar manner~\cite{watts,dodds}.

Among the several epidemic models, one of the simple contagion models is the so-called susceptible-infected-recovered (SIR) model~\cite{sir,sir_newman}, in which each node has one of three states, susceptible (denoted as $S$), infected ($I$), or recovered ($R$). Initially, all the nodes are in state $S$ except for one seed node in state $I$. The contagion process starts from a single node in state $I$. Each node in state $I$ transmits pathogens to its neighbors in state $S$ and infects each of them with probability $\kappa$; then, it changes its state to $R$ with unit probability. This contact process is repeated until the system reaches an absorbing state in which no infected node is left in the system. When the probability $\kappa$ is sufficiently small (large), the order parameter defined as the density of nodes in state $R$ after the system falls into the absorbing state, becomes $o(N)$ [$O(N)$]; i.e., the system falls into a subcritical (supercritical) state. In between, an epidemic transition occurs at $\kappa_c$, and the system exhibits critical behavior. It is known that when the dynamics starts from a single seed on Erd\H{o}s-R\'enyi (ER) random networks~\cite{ER}, the SIR model undergoes a continuous percolation transition following the universal behavior of ordinary percolation. 

The SIR model with multiple seeds has been considered~\cite{hasegawa2}, in which two percolation transitions occur successively at $\kappa_{c1}$ and $\kappa_{c2}$ as $\kappa$ is increased. The density of nodes in state $R$ is finite for $\kappa > \kappa_{c1}$, whereas the density of nodes in state $S$ disappears for $\kappa > \kappa_{c2}$. Thus, there exists a state of coexisting nodes in states $R$ and $S$ between $\kappa_{c1}$ and $\kappa_{c2}$.

The SIR model was extended to a two-step contagion model, in which a weakened state ($W$) can exist between the $S$ and $I$ states. Accordingly, this model is called the SWIR model~\cite{janssen,janssen_spinodal}. Nodes in state $W$ are involved in the reactions $S+I \to W+I$ and $W+I\to 2I$, which occurs in addition to the reactions $S+I\to 2I$ and $I\to R$ in the SIR model. The properties of the epidemic transition in the SWIR model were extensively investigated for the single-seed case~\cite{grassberger_pre,janssen, janssen_spinodal,chung, hasegawa1,choi_2016}. The order parameter defined as the density of nodes in state $R$ after an absorbing state is reached, displays a discontinuous transition, whereas other physical quantities such as the outbreak size distribution exhibit critical behaviors. Thus, the phase transition occurring in the SWIR model with a single seed is regarded as a mixed-order phase transition~\cite{choi_2016}.  The dynamic rule of the SWIR model is rather so simple that its underlying mechanism for the discontinuous behavior of the order parameter was disclosed~\cite{universal}. Moreover,  the mechanism turned out to be universal in other models such as $k$-core percolation~\cite{kcore1,kcore2,kcore3,baxter_prx}, the cascading failure model on interdependent networks~\cite{buldyrev, son, zhou, bk, review_mcc}, and the epidemic-related  models~\cite{watts,dodds,krapivsky, grassberger_pre,althouse_prl, chen, porter,grassberger_nphy,althouse_pnas}.

Here, we investigate the phase transitions of the SWIR  model with multiple seeds. The model with multiple seeds has been investigated in Refs.~\cite{hasegawa1, janssen_spinodal,hasegawa3}: The authors of Refs.~\cite{hasegawa1,hasegawa3} used the mean-field approach and performed numerical simulations, obtaining the phase diagram as a function of the reaction rates. The order parameter exhibits either a discontinuous or continuous transition depending on the density of the infectious seeds and mean degree of a given network~\cite{hasegawa1, hasegawa3}. In Ref.~\cite{janssen_spinodal}, the discontinuous transition is regarded as a spinodal transition, because there is no co-existence phase in the system while the order parameter jumps. Even though such results were obtained, the properties of the phase transitions and critical behaviors were not deeply  investigated yet. 

\begin{figure}[h]
\centering
\includegraphics[width=0.7\linewidth]{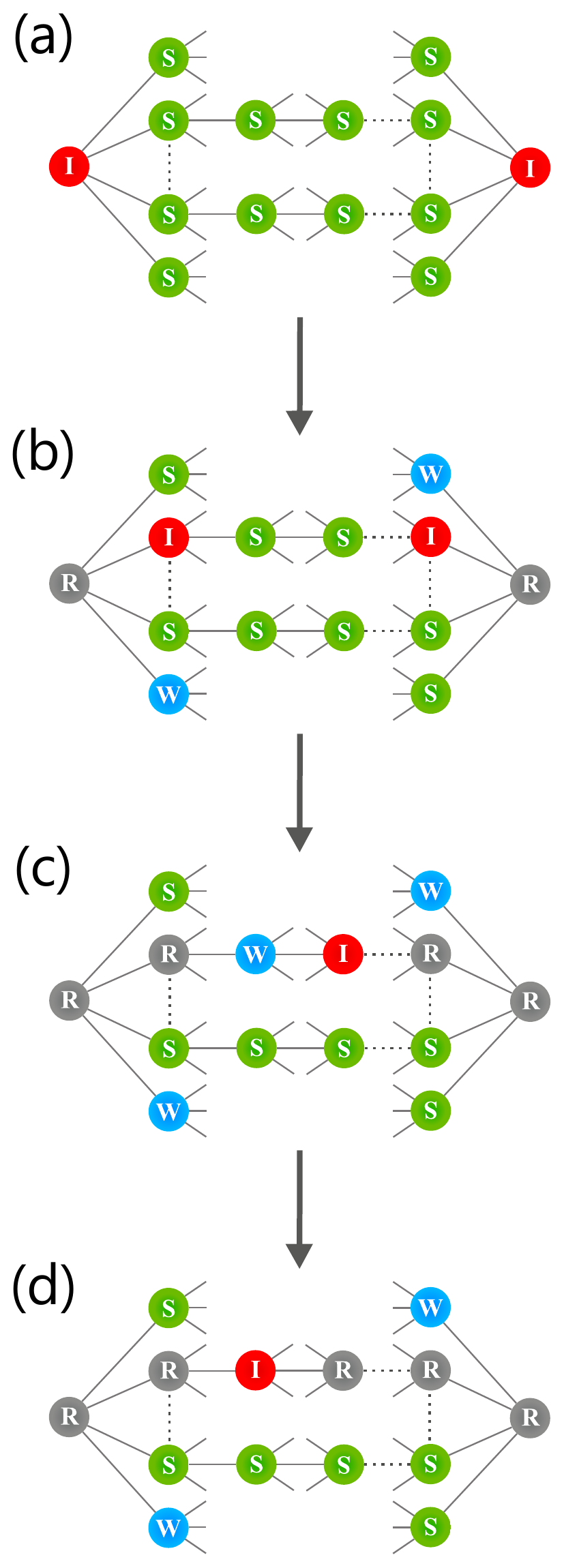}
\caption{Schematic illustration of epidemic spreading processes in the SWIR model with multiple seeds. (a) Epidemic spreading begins from each infectious node. (b) These nodes can infect susceptible neighbor nodes and change their state to either $I$ or $W$. (c) A node in state $I$ contacts a node in state $W$ from a different root. (d) Then, the node in state $I$ infects the node in state $W$ and changes its state to $I$. The two clusters merge.}
\label{fig1}
\end{figure}

Here, we reveal that the spread of contagion in the SWIR model with multiple seeds proceeds differently from that in the SWIR model with a single seed: in the multiple-seed case, the reactions $W+I\to 2I$ often occur even in early time steps, because nodes in states $W$ and $I$ involved in that reaction can originate from different seeds (see Fig.~\ref{fig1}). We note that the number of multiple seeds was taken as $O(N)$. On the contrary, in the single-seed case, such reactions rarely occur until the system reaches a characteristic dynamic step $n_c(N)\sim N^{1/3}$: When  dynamic step $n$ is less than $n_c(N)$, the reactions $S+I\to 2I$ and $I\to R$ are dominant but the number of nodes in $R$ still remains as $o(N)$. The contagion spreads in the form of a branching tree. When the dynamics reaches $n_c(N)$,  the branching tree forms long-range loops due to finite-size effect. Once such loops form, the reaction $W+I\to 2I$ occurs massively, in which the nodes in state $W$ were generated in early time steps. Thus, the density of nodes in state $R$ increases drastically as many as $O(N)$ in short time steps. Due to these different contagion mechanisms, the properties of epidemic transitions in the multiple seed case become different from those in the single seed case. We will determine the full set of critical exponents describing the phase transitions in the multiple seed case, and compare them with those obtained in the single seed case~\cite{choi_2016}.    

This paper is organized as follows: In Sec. II, we present the rules of the SWIR model in detail. In Sec. III, we set up the self-consistency equation to derive the mean-field solution using the local tree approximation of the order parameter for the epidemic transition on the ER networks. We show that, depending on the initial density of infectious nodes, different types of phase transition can occur. In Sec. IV, we report numerical results for the epidemic transitions. In the final section, a summary and discussion are presented. 

\section{The SWIR model}

The SWIR model with multiple seeds is simulated on ER networks with $N$ nodes. Initially, $N\ro$ nodes are selected randomly from among those $N$ nodes and assigned to state $I$; the other nodes are assigned to state $S$. At each time step $n$, the following processes are performed. (i) All the nodes in state $I$ are listed in random order. (ii) The states of the neighbors of each node in the list are updated sequentially as follows: If a neighbor is in state $S$, it changes its state in one of the two ways: either to $I$ with probability $\kappa$ or to $W$ with probability $\mu$. If a neighbor is in the state $W$, it changes to $I$ with probability $\eta$, where $\kappa$, $\mu$, and $\eta$ are the contagion probabilities for the respective reactions. (iii) All nodes in the list change their states to $R$. This completes the time step, and we repeat the above processes until the system reaches an absorbing state in which no infectious node is left in the system. The reactions are summarized as follows:
\begin{eqnarray}
\rm{S+I}  &\buildrel{\kappa}\over \longrightarrow& {\rm I+I}, \label{si_ii} \\
\rm{S+I} &\buildrel{\mu}\over \longrightarrow& \rm{W+I}, \label{si_wi} \\
\rm{W+I}  &\buildrel{\eta}\over \longrightarrow& \rm{I+I}, \label{wi_ii} \\
\rm{I}  &\buildrel{1}\over \longrightarrow& \rm{R}. \label{i_r} 	
\end{eqnarray}

The order parameter exhibits a discontinuous transition at a transition point $\kappa_c$ when $\rho_0$ is less than a critical value $\rho_c$, 
and it shows a continuous transition at $\kappa_c$ when $\rho_0=\rho_c$ for given parameter values $z$, $\mu$, and $\eta$, where $z$ is the mean degree of a given ER network. 

\section{Self-consistency equation and physical solutions}

In an absorbing state, each node is in one of three states: the susceptible $S$, weakened $W$, and recovered $R$ states. The order parameter $m(\kappa)$, the density of nodes in state $R$ in an absorbing state, is written using the local tree approximation as 
\begin{equation}\label{rho_r}
m(\kappa) = \rho_0 + (1-\rho_0)\sum_{k=1}^{\infty}P_d(k) \sum_{\ell=1}^{k} \binom{k}{\ell}q^{\ell}(1-q)^{k-\ell} P_R (\ell).
\end{equation}
The first term in Eq.~(\ref{rho_r}), $\ro $, is the initial density of infected nodes. In the second term, the factor $(1-\ro)$ represents the probability that a node is originally in state $S$. $P_{d}(k)$ is the probability that a randomly selected node has degree $k$; $q$ is the probability that an arbitrarily chosen edge leads to a node that is in state $R$ but not infected through the chosen edge in the absorbing state. Thus, $P_d(k)\binom{k}{\ell}q^{\ell}(1-q)^{k-\ell}$ is the probability that a node has degree $k$ and $ \ell $ of them are in state $R$ in the absorbing state. $P_R(\ell)$ is the conditional probability that a node is finally in state $R$, provided that it was originally in state $S$ and its $\ell$ neighbors are in state $R$ in the absorbing state.

Similarly to $P_R(\ell) $, we define $P_S(\ell)$ as the conditional probability that a node remains in state $S$ in the absorbing state, provided that it has $\ell$ neighbors in state $R$ and was originally in state $S$. $P_W(\ell)$ is defined similarly. We note that for a certain node to have $\ell $ neighbors in state $R$ in the absorbing state means that the node receives $ \ell $ attempts to infect it when the recovered neighbors are in state $I$. Thus, a node still remaining in state $S$ with $\ell$ neighbors in state $R$ has to be unchanged from $\ell$ infection attempts through the entire process. Thus, we obtain
\begin{equation}
\ps(\ell) = (1-\kappa-\mu)^{\ell}.
\label{p_s}
\end{equation}
Next, the probability $\pw(\ell)$ is given as
\begin{equation}
\pw(\ell) = \sum_{j=0}^{\ell-1} (1-\kappa-\mu)^{j} \mu (1-\eta)^{\ell-j-1},
\label{p_w}
\end{equation}
where $j$ denotes the number of attacks that a node sustains before it changes to state $W$. Using the relation $\ps(\ell)+\pw(\ell)+\pr(\ell)=1$, one can determine $\pr(\ell)$ in terms of $\ps$ and $\pw$. 

The local tree approximation allows us to define $q_n$ similarly to $q$ but now at the tree level $n$. 
 The probability $q_{n+1}$ can be derived from $q_n$ as follows:
\begin{widetext}
\begin{equation} \label{eq:sce}
q_{n+1}=\rho_0+(1-\rho_0)\sum_{k=1}^{\infty} \frac{kP_d(k)}{z}\sum_{\ell=1}^{k-1} \binom{k-1}{\ell} q_n^{\ell}(1-q_n)^{k-1-\ell} \pr(\ell) \equiv \rho_0+(1-\rho_0)f(q_n),
\end{equation}
\end{widetext}
where the factor $kP_d(k)/z$ is the probability that a node connected to a randomly chosen edge has degree $k$. As a particular case, when the network is an ER network with $P_d(k)=z^k e^{z}/k!$, $f(q_n)$ becomes
\begin{equation} 
f(q_n)=1-\Big(1+\dfrac{\mu}{\eta-\kappa-\mu}\Big)e^{-(\kappa+\mu)q_n z}+\dfrac{\mu}{\eta-\kappa-\mu}e^{-\eta q_n z}. \label{eq:f_q}
\end{equation}
 
Eq.~(\ref{eq:sce}) reduces to a self-consistency equation for $q$ for given epidemic parameter values in the limit $n\to \infty$. Once we obtain the solution of $q$, we can obtain the outbreak size $m(\kappa)$ using Eq.~(\ref{rho_r}). For ER networks, however, $m(\kappa)$ becomes equivalent to $q$, thus the solution of the \self equation Eq.~(\ref{eq:sce}) yields the order parameter. We remark that the methodology we used here is similar to those used in previous studies of epidemic spreading on complex networks~\cite{ dodds, chung, janssen_spinodal,grassberger_pre,hasegawa1}.

Hereafter, we set $ \mu=\kappa $ for convenience and define a function 
\begin{equation}\label{g_m_def}
F(m,\ro)\equiv f(m)-\frac{m}{1-\rho_0}+\frac{\rho_0}{1-\rho_0}.
\end{equation} 
Using formula (\ref{eq:f_q}), we approximate $F(m,\ro)$ in the limit $m\to 0$ as 
\begin{equation} \label{eq:G_q}
F(m,\ro)=\frac{\ro}{1-\ro}+a m+b m^2+ c m^3+O(m^4), 
\end{equation}
where 
\begin{eqnarray}
a&=&\kappa z-(1/(1-\rho_0)), \label{a_c} \\
b&=&\frac{1}{2}\kappa(\eta-2\kappa)z^2, \label{b_c}\\
c&=&\frac{1}{6}\kappa(4\kappa^2-2\eta\kappa-\eta^2)z^3. \label{c_c} 
\end{eqnarray}
For convenience, we neglect the higher-order terms and redefine $F(m,\ro)$ as
\begin{equation}
F(m,\ro)\equiv \frac{\ro}{1-\ro}+am+bm^2+cm^3.
\label{eq:h_q}
\end{equation} 

Depending on the relative magnitudes of $a$ and $b$, various solutions of the self-consistency equation $F(m,\ro)=0$ can be obtained. However, we need to check whether these solutions are indeed physically relevant in the steady state when we start epidemic dynamics from a certain initial condition. The stability criterion was established in a previous work~\cite{choi_2016}: The solution $F(m_0 ,\ro) =0$ is stable if and only if $\partial F(m,\ro)/\partial m |_{m=m_0} < 0 $.

\begin{figure}[H]
\centering
\includegraphics[width=0.9\linewidth]{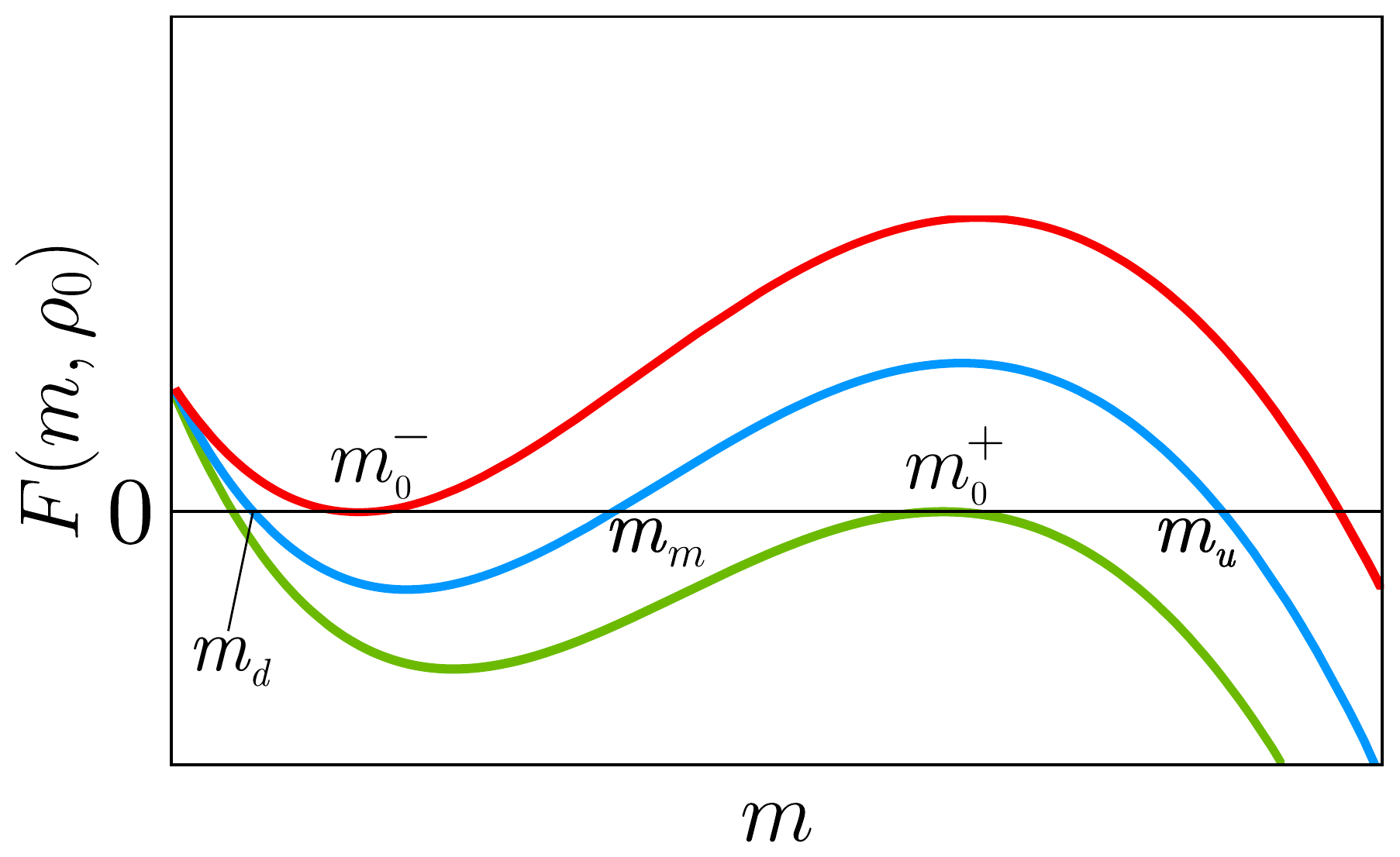}
\caption{Schematic plot of $F(m,\ro)$ versus $m$ for $0<\ro <\rho_c$. Curves represent $F(m,\ro)$ for different $\kappa $. $m_d$, $m_m$, and $m_u$ are the solutions of $F(m,\ro)=0$, where $ m_{d} < m_m < m_u $. $m_0^{\pm}$ are the solutions of $F(m,\ro)=dF(m,\ro)/dm=0$ with $m_0^- < m_0^+$.
}
\label{fig2}
\end{figure}

\begin{figure}[H]
\centering
\includegraphics[width=0.9\linewidth]{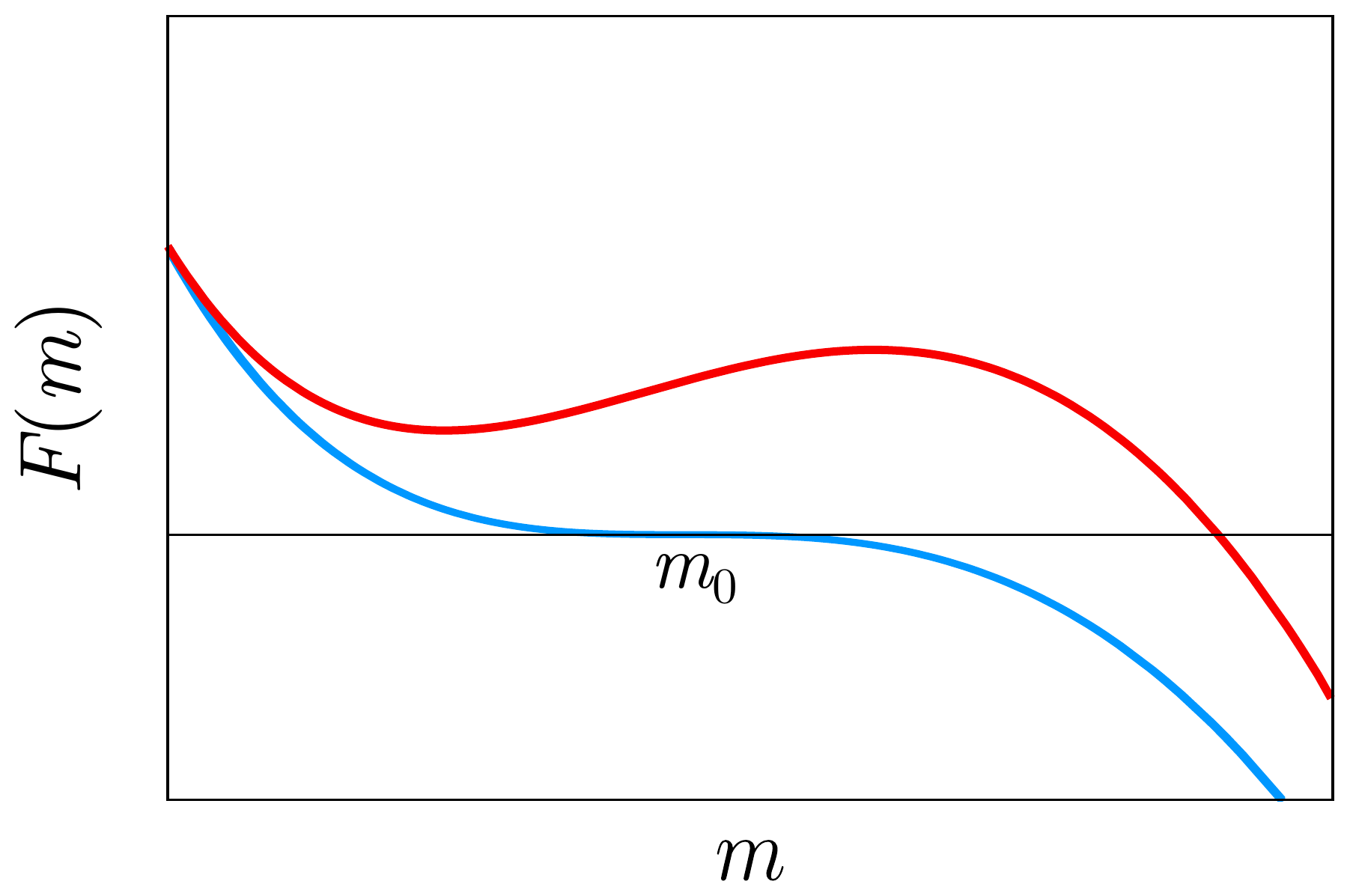}
\caption{Schematic plot of $F(m,\rho_c)$ versus $m$ for $\ro=\rho_c$. There exists a solution $m_0$ at which the \self equations $F(m_0,\rho_c)=0$ and $d^2F(m,\rho_c)/d^2m|_{m=m_0}=0$ hold.}
\label{fig3}
\end{figure}

The equation of state in the steady state can be obtained using $F(m,\ro)=0$. We notice that $F(m=0,\ro)=\ro/(1-\ro) > 0$ and $F(m=\infty,\ro)=-\infty$ because $c < 0$, as shown in Fig.~\ref{fig2}. We examine the solutions of $dF(m,\ro)/dm=0$, which are obtained as  
\begin{equation} \label{eq:hq}
m^{\pm}=-\dfrac{b}{3c} \pm \sqrt{ \dfrac{b^2}{9c^2}-\dfrac{a}{3c}},
\end{equation}
where $a$, $b$, and $c$ are given in formulas (\ref{a_c})--(\ref{c_c}). Note that $a$ depends on $\ro$. At these extreme points $ m^{\pm} $, $F(m,\ro)$ has either a local maximum or a local minimum. For a given $\ro$, $z$, and $\eta$, both $ m^{\pm} $ values exist, and they are positive in the range of $ \kappa_d <\kappa < \kappa_a $, where $ b^2 / 9c^2 - a/3c =0 $ at $ \kappa=\kappa_d $, and $ a=0 $ at $ \kappa=\kappa_a $. For a given $ z $ and $ \eta $, diverse types of phase transitions occur depending on $\ro$. When $ \ro $ is less than a certain value $ \rho_c $, the order parameter jumps at a transition point. On the other hand, when $ \ro \geq \rho_c $, the order parameter increases continuously with $ \kappa $. At $ \ro=\rho_c $, $ m^+=m^-=m_0$ and $ F(m_0,\rho_c)=0 $ at $ \kappa=\kappa_d=\kappa_c $, as schematically shown in the blue (lower) curve in Fig.~\ref{fig3}.

\subsection{When $\ro < \rho_c$}  

\begin{figure}[H]
\centering
\includegraphics[width=0.9\linewidth]{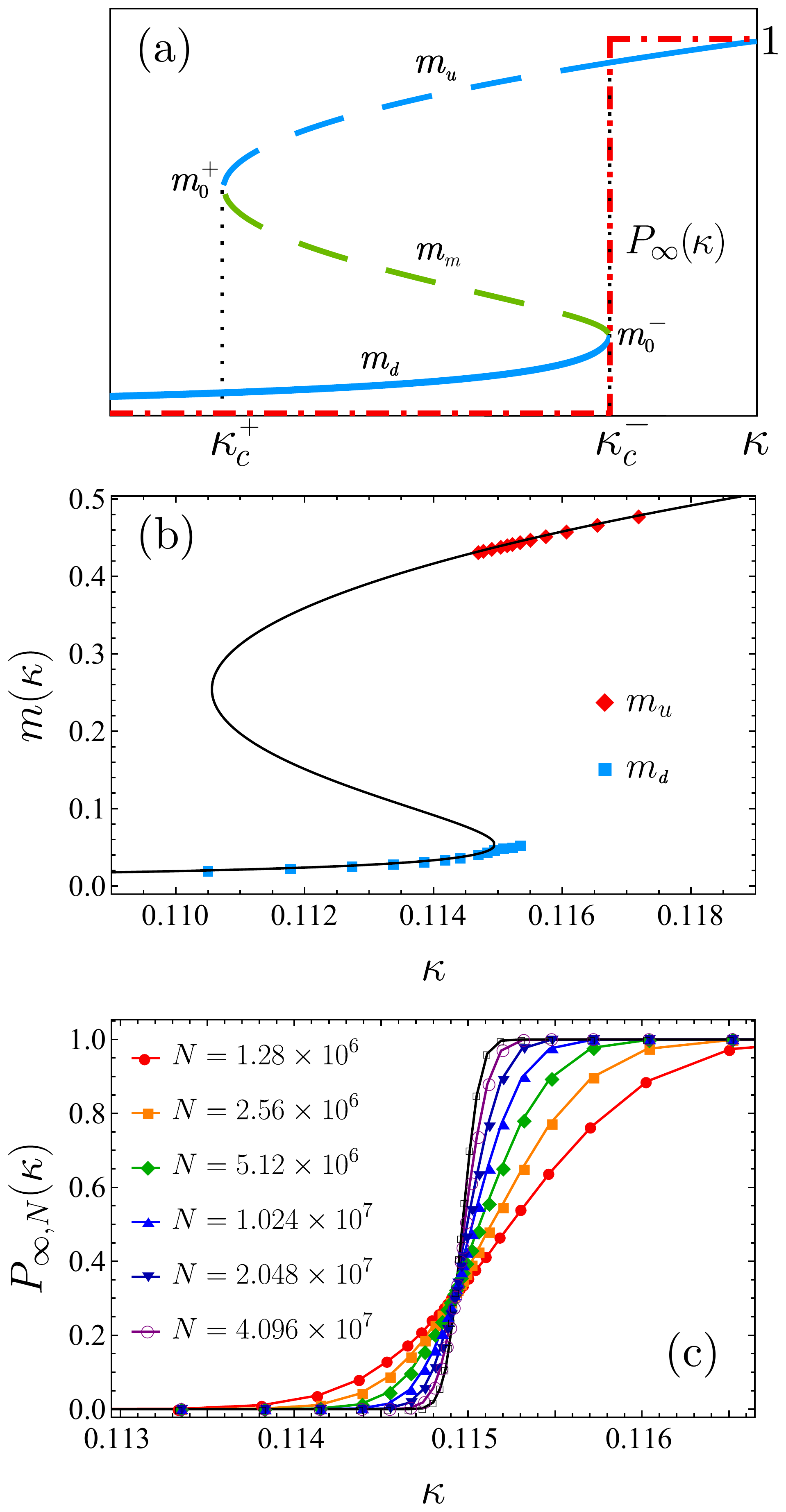}
\caption{(a) Schematic plot of the order parameter $m(\kappa)$ versus $\kappa$ for $\ro<\rho_c$. Thick solid (dashed) curves represent stable (unstable) solutions of the \self equation. Dashed-dotted lines represent the pandemic probability $P_{\infty}(\kappa)$. (b) Plot of $m(\kappa)$ versus $\kappa$ for ER networks with mean degree $z=8 $ and $ \ro=2 \times 10^{-3} $. Solid curve represents analytic solution of the self-consistency equation. Red dots (blue squares) represent averaged values of $m_u $ ($ m_d $) obtained by numerical simulations on ER networks of system size $ N=4.096\times 10^7 $. (c) Plot of $ P_{\infty,N} $ versus $\kappa$ for various system sizes $N$. Data are obtained for ER networks with mean degree $z=8 $ and $ \ro=2\times 10^{-3}$. At $\kappa=\kappa_c^- \approx 0.11495$, $ dP_{\infty,N}/d\kappa \sim N^{1/2} $, which implies that $ P_{\infty}(\kappa)$ behaves like a step function in the limit $ N \rightarrow \infty $, as depicted schematically in (a) with dashed-dotted lines.}
\label{fig4}
\end{figure}

When $ \ro<\rho_c,$ there exists a range of $ \kappa $ in which $F(m,\ro)=0$ has more than one solution, as shown in Fig.~\ref{fig2}. The order parameter $m$ versus $\kappa $ is shown in Fig.~\ref{fig4}(a) and (b). 
In particular, when $\kappa$ has a certain value $\kappa_c^-$, $m^-$ obtained using Eq.~(\ref{eq:hq}) satisfies $F(m^-,\ro)=0$. The $m^-$ value at $\kappa_c^-$ is denoted as $m_0^-$. We also define $\kappa_c^+$ and $m_0^+$ similarly to $m^+$ in Eq.~(\ref{eq:hq}). 
We note that $\kappa_c^+ < \kappa_c^-$. Depending on the magnitude of the reaction probability $\kappa$ relative to $\kappa_c^+$ and $\kappa_c^-$, the order parameter behaves differently, as follows:

i) For $\kappa < \kappa_c^+$, there exists one stable solution $m=m_d(\kappa)$, which increases slowly with $\kappa$. It is obtained that 
${m_d}\approx\ro/(1-\kappa z)+O(\ro^2)$.

ii) At $\kappa=\kappa_c^+$, there exist two solutions, $m_d$ and $m_0^+$ ($m_d < m_0^+$). However, $m_0^+$ is not accessible because $m_d$ is stable.  
  
iii) When $\kappa_c^+ < \kappa < \kappa_c^-$, there exist three solutions, $m_d$, $m_m$, and $m_u$, with relative magnitudes $m_d < m_m < m_u$; however, the solution $m_m$ is unstable. Thus, only $m_d$ is accessible from the initial density $\rho_0 < m_d$. The order parameter behaves as $m_0^--m_d(\kappa) \sim(\kappa_c^- -\kappa )^{1/2}$ for $\kappa < \kappa_c^-$. Thus, the critical exponent of the order parameter is obtained as $\beta=1/2$.

iv) At $\kappa=\kappa_c^-$, there exist two stable solutions, $m_0^-$ and $m_u$. Thus, the order parameter jumps between the two values, exhibiting discontinuous behavior. Hence, a hybrid phase transition occurs at the point $(\kappa_c^-, m_0^-)$. 

v) For $\kappa > \kappa_c^-$, there exists one solution, denoted as $m_u(\kappa)$, which increases with $\kappa$ as $m_u(\kappa)-m_u(\kappa_c^-)\sim (\kappa-\kappa_c^-)$.  

\subsection{When $\ro=\rho_c$} 

\begin{figure}[H]
\centering
\includegraphics[width=0.9\linewidth]{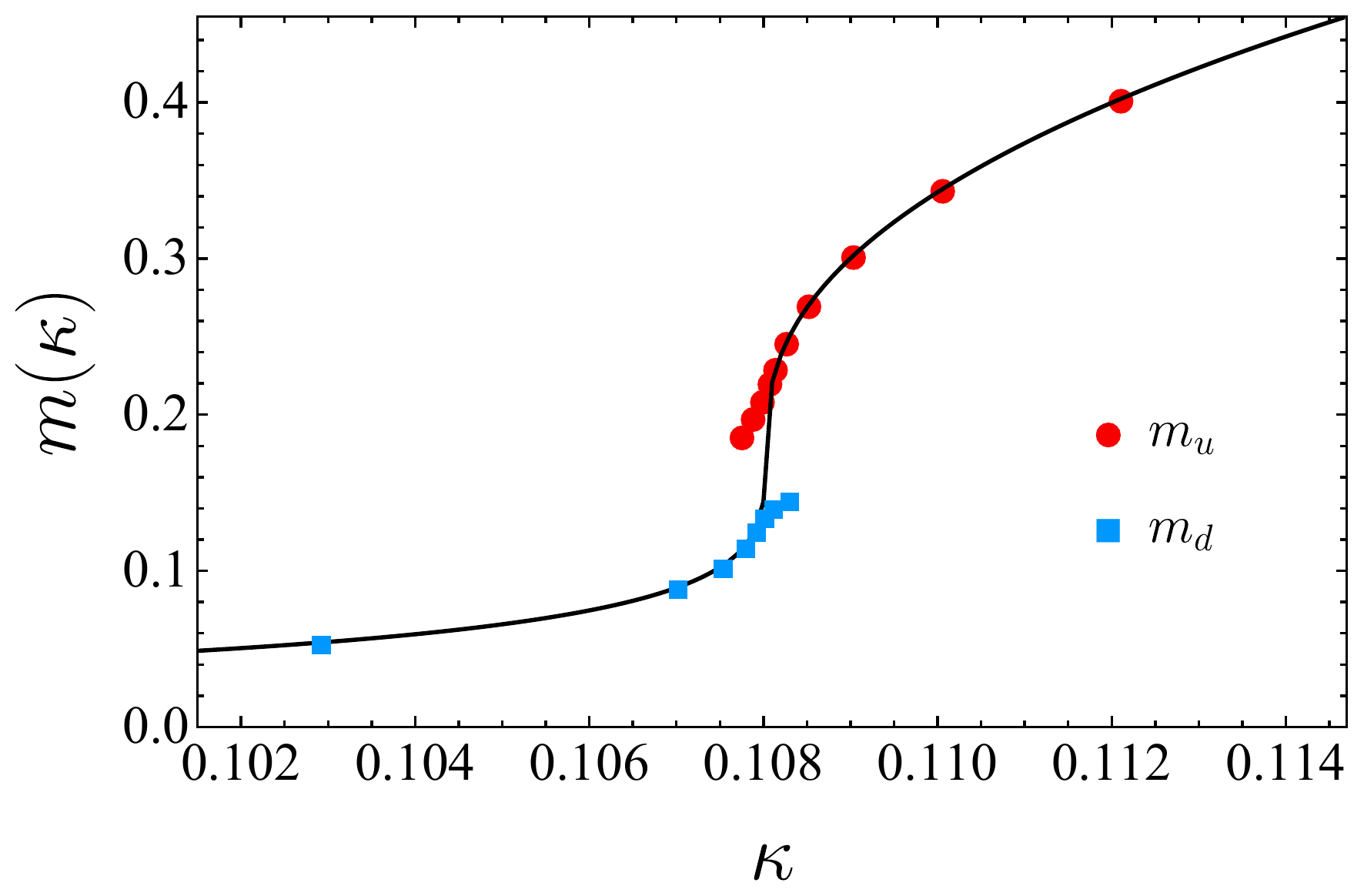}
\caption{Plot of $m(\kappa)$ versus $\kappa$  for ER networks with mean degree $z=8 $ and $ \ro=\rho_c\approx 0.00747762 $. Solid curve represents analytic solution of the self-consistency equation. Red dots (blue squares) represent average values of $m_u $ ($ m_d $) obtained by numerical simulations on ER networks of the system size $ N=1.024\times 10^7 $.}
\label{fig5}
\end{figure}

When $\ro=\rho_c$, there exists a reaction probability $\kappa_c$ that satisfies the relation $F(m_0,\rho_c)=dF(m,\rho_c)/dm|_{m=m_0}=0$, and $b^2-3ac=0$. Thus, the two solutions, $m_0^-$ and $m_0^+$, reduce to the same one, which is denoted as $m_0$. The function $F(m,\rho_c)$ versus $m$ is shown in Fig.~\ref{fig3}, and the order parameter $m$ versus $\kappa$ is shown with the analytic solution and simulation data in Fig.~\ref{fig5}. At $\kappa_c$, singular behavior occurs, and the order parameter $m$ behaves as $m-m_0\sim |\kappa-\kappa_c|^{1/3}$ on both sides. The derivation of this exponent $1/3$ is presented in the Appendix.

\section{Numerical results}

To estimate various critical exponents, we perform extensive numerical simulations on ER networks with mean degree $z=8$. For simplicity, the reaction probability $\mu$ is set equal to $\kappa$, and $\eta=1/2$. With these parameter values, we determine $\rho_c$ as precisely as $0.00747762$, which we will use in numerical analysis later. For $\ro < \rho_c$, we take $\ro=2 \times 10^{-3}$ in the simulations. We take the average over 50 different dynamics samples for each of 1,600--4,000 network configurations. Thus, 80,000--200,000 configuration averages were taken to obtain each data point.

\subsection{When $\ro<\rho_c$} 

\begin{figure}[h]
\centering
\includegraphics[width=0.9\linewidth]{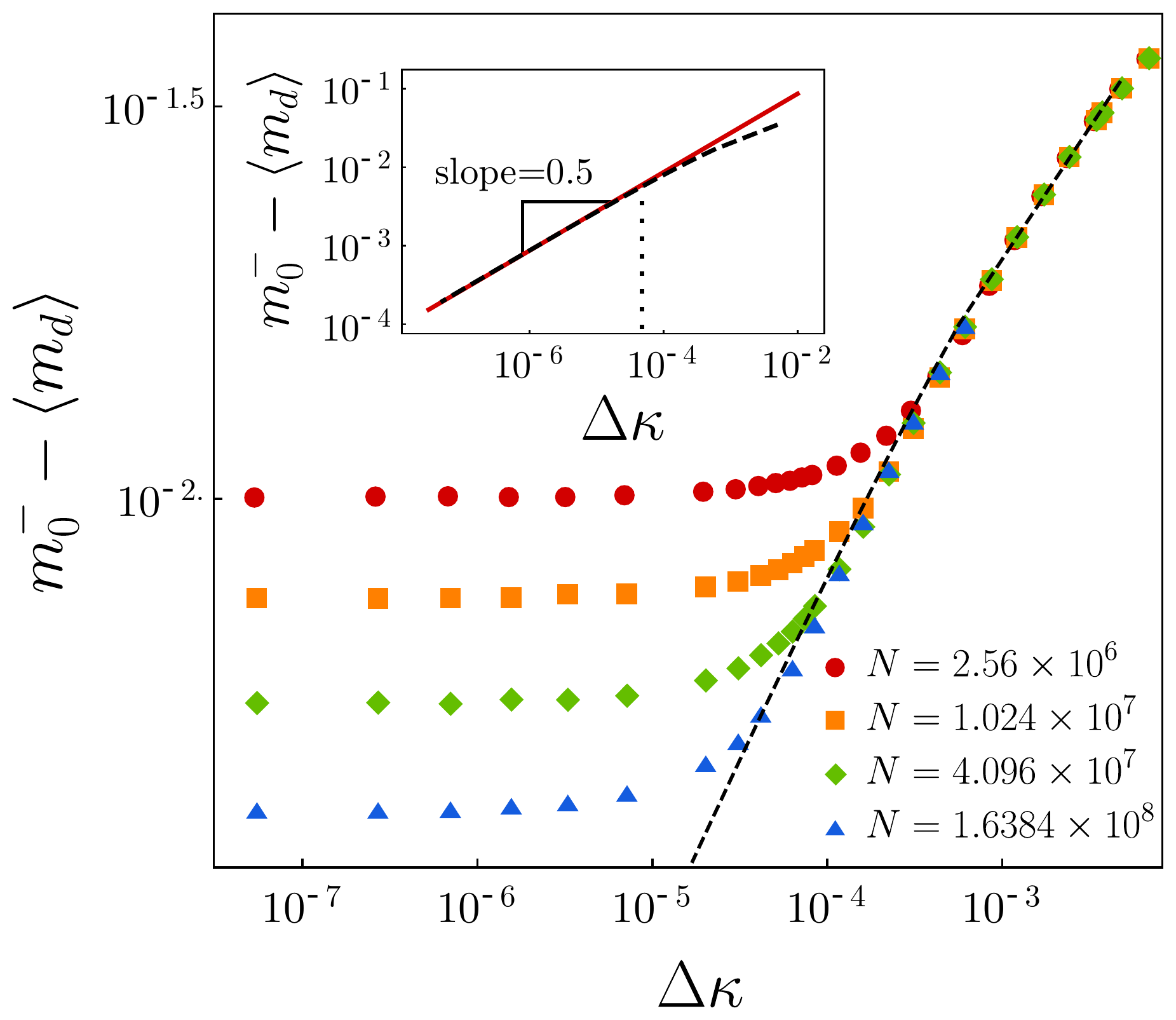}
\caption{Plot of $m_0^--m_d$ versus $\Delta \kappa=\kappa_c^- -\kappa$ in a double logarithmic scale. Data are obtained for ER networks with mean degree $z=8$ with $\rho_0 =2\times 10^{-3}$. The black dashed curve represents the analytical solution. For $ N=1.6384\times 10^8 $ ($\triangle$), crossover behavior is likely to occur at $\Delta\kappa \approx 10^{-4.2}$, which is roughly close to the point from which the analytical solution (black dashed curve in the inset) of $m_0^- - m_d$ deviates from the straight line with a slope of 0.5. 
} 
\label{fig6}
\end{figure}

We found analytically that the order parameter behaves as $m_0^-- m_d(\kappa)\sim (\Delta \kappa)^{\beta}$ with $\beta=1/2$ in the thermodynamic limit, where $ \Delta\kappa \equiv \kappa_c^--\kappa $. The main panel of Fig.~\ref{fig6} shows $ m_0^--\langle m_d(\kappa) \rangle $ versus $\Delta\kappa \equiv \kappa_c^--\kappa$ in a double logarithmic scale. Data points in the figure are obtained numerically from systems of several selected sizes $N$, and the dashed line is obtained from the analytic solution of Eq.~(\ref{eq:sce}) by taking the limit $n\to \infty$, which is valid in the thermodynamic limit. We find that the data points saturate to constant values asymptotically as $\Delta \kappa\to 0$, whereas they overlap with the dashed curve as $\Delta \kappa$ is increased. As shown in the inset, also in a double logarithmic scale, the dashed curve follows the line with a slope of $0.5$ in the region $\Delta \kappa < \Delta \kappa^* \approx 10^{-4.2}$; however, it deviates from the line in the opposite region beyond $\Delta \kappa^*$. This fact implies that conventional finite-size scaling analysis is valid for systems with size larger than $O(10^8)$. However, it would be impractical to perform simulations with such huge system sizes.    

\begin{figure}[h]
\centering
\includegraphics[width=0.9\linewidth]{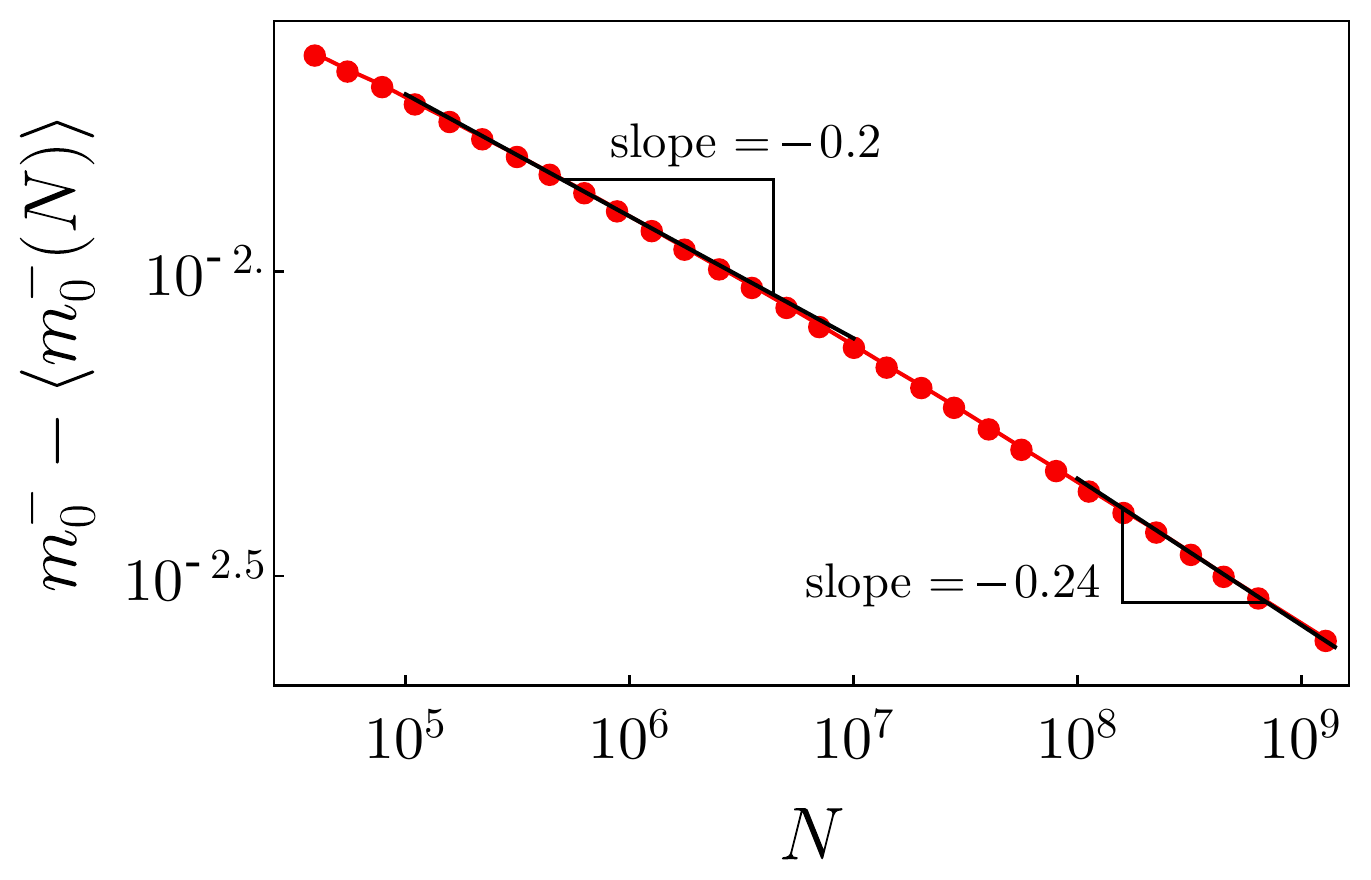}
\caption{Plot of $m_0^--\langle m_0^-(N) \rangle$ versus $N$ at $\kappa=\kappa_c^-$ in a double logarithmic scale. Data are obtained for ER networks with mean degree $z=8$. $\rho_0 =2\times 10^{-3}$ is used. The slope of the data point corresponds to $\beta/\nubar$. As the system size is increased, crossover behavior appears in the slope from $-0.2$ to $-0.24$, which indicates that $ \beta/\nubar \approx 0.24 $ in the limit $N\to \infty$.} 
\label{fig7}
\end{figure}

Following the conventional finite-size scaling theory, 
\begin{equation} 
m_0^-(\infty) - \langle m_0^-(N) \rangle \sim N^{-\beta / \nubar}
\label{finite_m}
\end{equation} 
at $\kappa_c^-$. We check this relation in Fig.~\ref{fig7}. For small system sizes $N$, $ \beta/\bar{\nu} $ seems to be about 0.2, whereas it is estimated to be $\approx 0.24$ for large $N$. Again the crossover occurs between the system sizes $N=10^7$ and $10^8$. We could obtain a more reliable value for the exponent ratio $\beta/\nubar$ for somewhat larger system sizes, but that is impractical.  

\begin{figure}[H]
\centering
\includegraphics[width=0.9\linewidth]{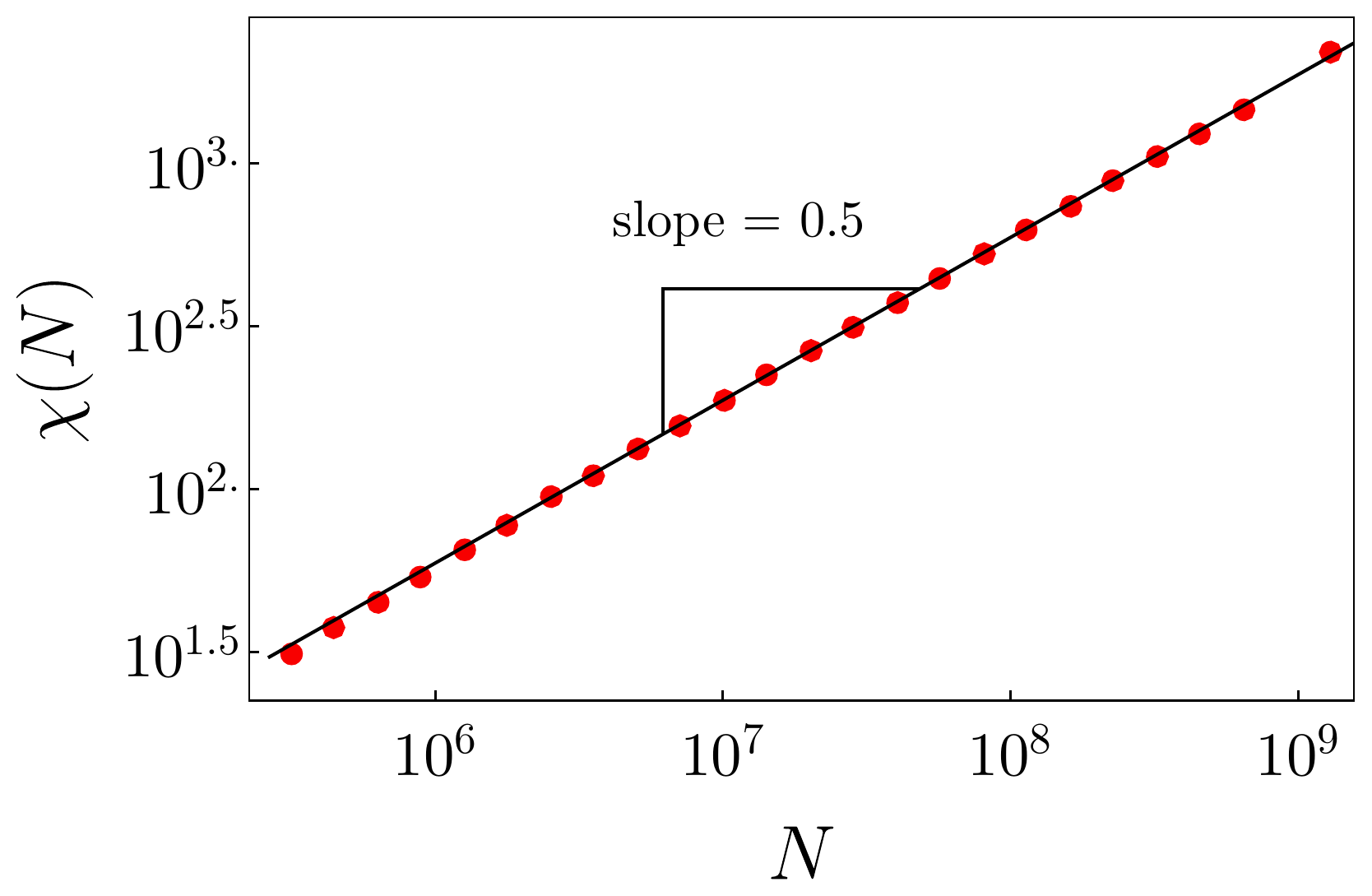}
\caption{Plot of the susceptibility $\chi$ versus $N$ at $ \kappa=\kappa_c^- $ in a double logarithmic scale. Data are obtained for ER networks with mean degree $z=8$. $\ro =2\times 10^{-3}$. Here, the slope of the data points corresponds to $\gamma / \nubar$, which is estimated to be $\approx 0.5$.}
\label{fig8}
\end{figure}

The fluctuation of the order parameter $\chi(\kappa)\equiv N(\langle m_d^2 \rangle - \langle m_d \rangle^2)$ diverges as $\sim (\kappa_c^--\kappa)^{-\gamma}$. For finite systems of size $ N $, it is expected that $ \chi \sim N^{\gamma/\nubar}$ at $\kappa=\kappa_c^-$. From the simulation data, we obtain $ \gamma / \nubar \approx 0.5 $, as shown in Fig.~\ref{fig8}. 

With the measured values $\beta/\nubar\approx 0.24$ and $\gamma/\nubar\approx 0.5$ and the analytic result $\beta=1/2$, we guess $ \nubar=2$ and then $\gamma=1$. If we use those values, then the hyperscaling relation $2\beta + \gamma=\bar{\nu}$ would hold.

\subsection{When $\ro=\rho_c$}

\begin{figure}[h]
\centering
\includegraphics[width=0.9\linewidth]{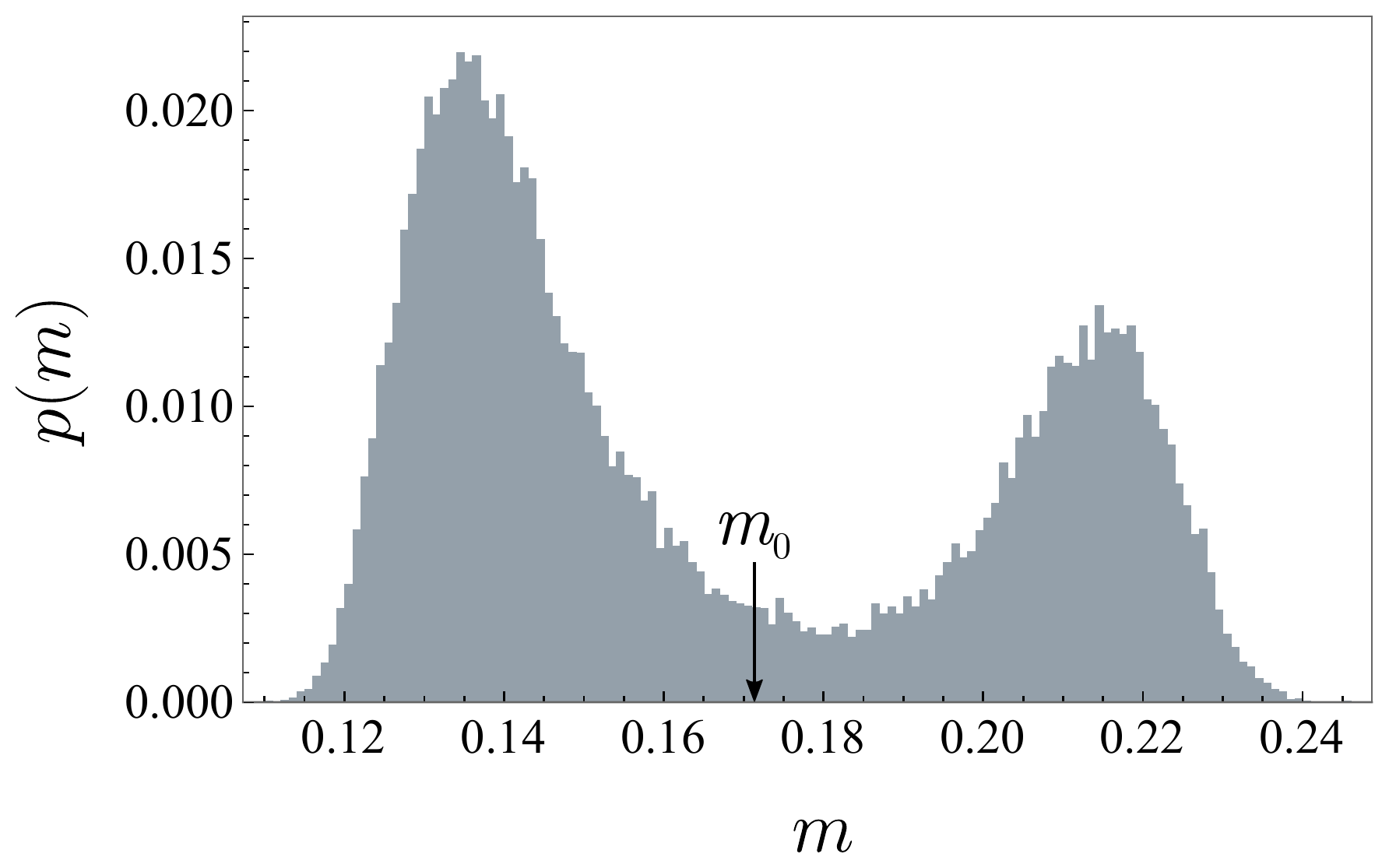}
\caption{Plot of the histogram of the order parameter $p(m) $ at $ \kappa_c\approx0.108021$. Data are obtained for ER networks of $ N=2.048 \times 10^7 $ with mean degree $z=8 $ and $\ro=\rho_c\approx 0.00747762 $. Even though simulations were performed at $\kappa_c$ and $\rho_c$, the distribution of the order parameter exhibits two peaks, a prototypical pattern of a discontinuous transition due to the finite size effect. As $N$ is increased, we expect that the two peaks converge and become a single peak.}
\label{fig9}
\end{figure}

\begin{figure}[H]
\centering
\includegraphics[width=0.9\linewidth]{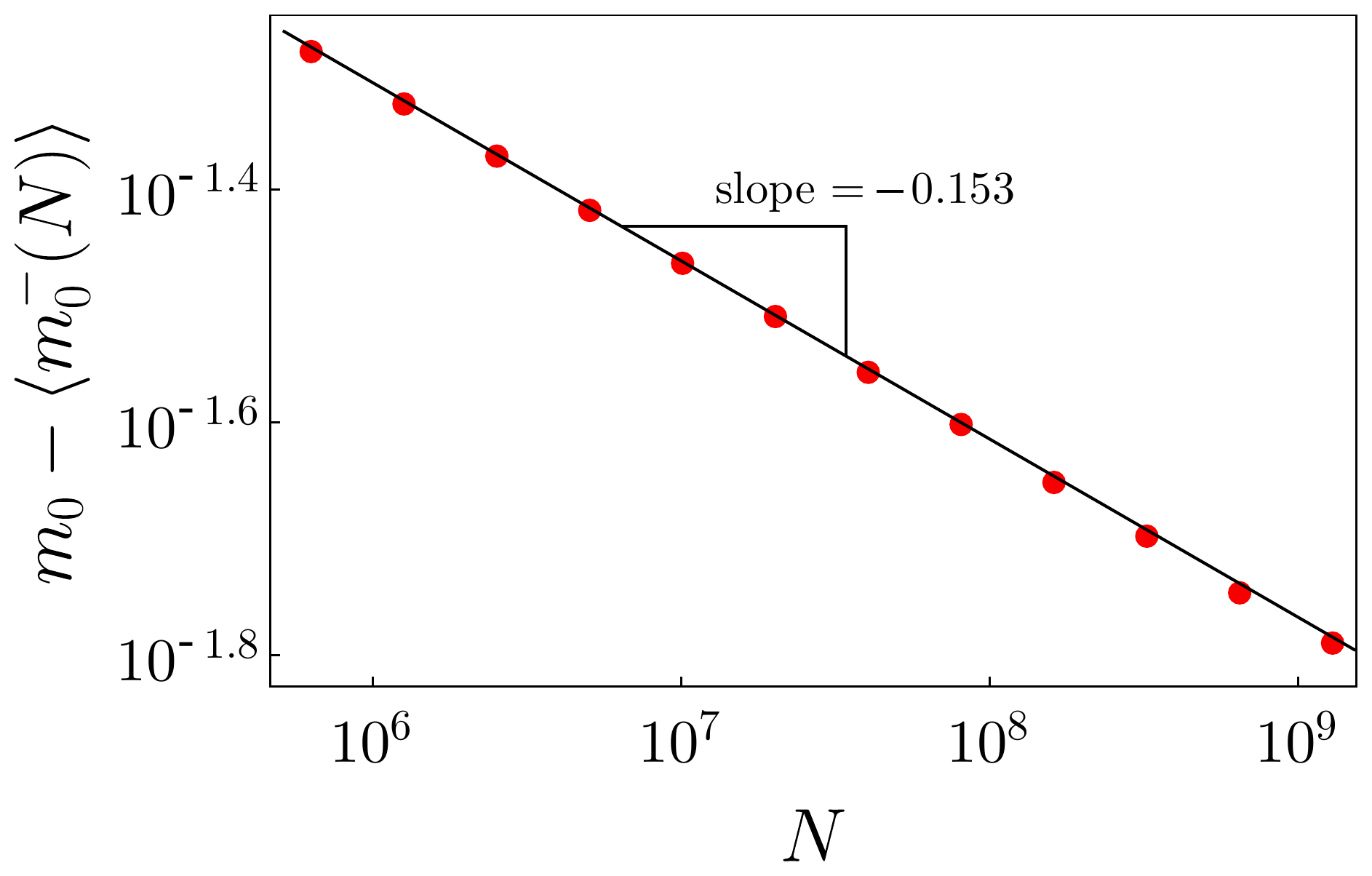}
\caption{Plot of $m_0-\langle m_0^-(N)\rangle$ versus $N$ at $\kappa_c\approx 0.108021$. Data are obtained for ER networks with mean degree $z=8$. $\ro=\rho_c$ is taken as $\approx 0.00747762$. $\beta/\nubar$ is estimated to be $\approx 0.153$.}
\label{fig10}
\end{figure}

\begin{figure}[H]
\centering
\includegraphics[width=0.9\linewidth]{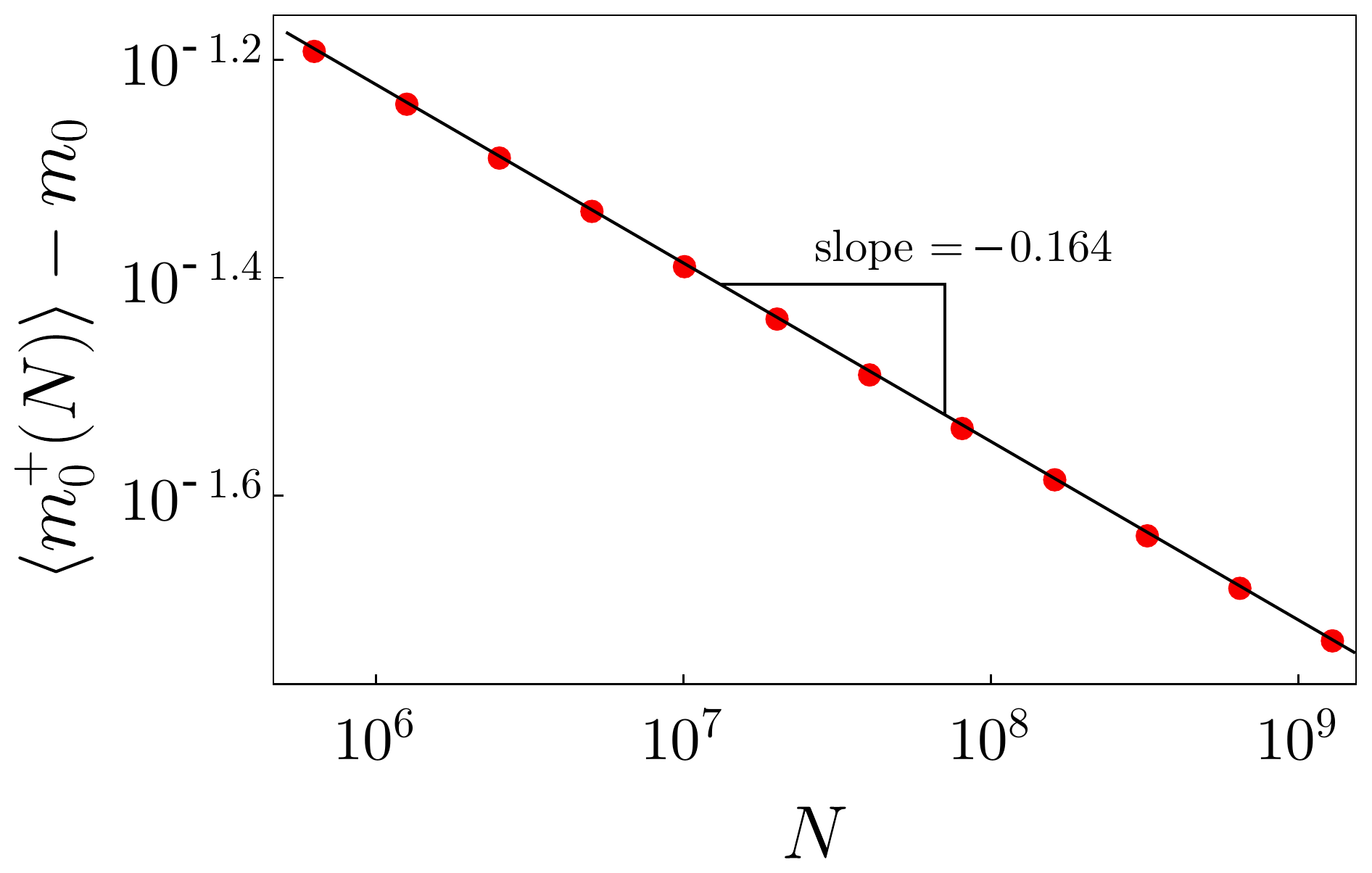}
\caption{Plot of $\langle m_0^+(N)\rangle-m_0$ versus $N$ at $\kappa_c \approx 0.108021$. Data are obtained for ER networks with mean degree $z=8 $. $\ro=\rho_c$ is taken as $\approx 0.00747762$. $\beta^{\prime}/\nubar^{\prime}$ is estimated to be $\approx 0.164$.}
\label{fig11}
\end{figure}

At $ \ro=\rho_c $, the jump in the order parameter does not appear, and $ m_0^+ = m_0^- $ at $ \kappa=\kappa_c $ in the thermodynamic limit. In finite systems, however, the order parameter can still exhibit a jump in some samples. Thus, the order parameter distribution $p(m)$ accumulated over different samples exhibits two separate peaks, as shown in Fig.~\ref{fig9}. We regard the data points of $p(m)$ in the region $m < m_0$ ($m > m_0$), where $m_0$ has the theoretical value $0.171405\dots$, as those obtained from $m_0^-(N)$ ($m_0^+(N)$) for different samples. At $ \kappa=\kappa_c$, in finite systems, we obtain the power-law behaviors $m_0-\langle m_0^-(N)\rangle \sim N^{-\beta / \nubar}$ with $ \beta / \nubar \approx 0.153 $ (Fig.~\ref{fig10}) and $ \langle m_0^+(N)\rangle - m_0 \sim N^{-\beta^{\prime}/ \nubar^{\prime}} $ with $\beta^{\prime}/\nubar^{\prime}\approx 0.164 $ (Fig.~\ref{fig11}). 

\begin{figure}[H]
\centering
\includegraphics[width=0.9\linewidth]{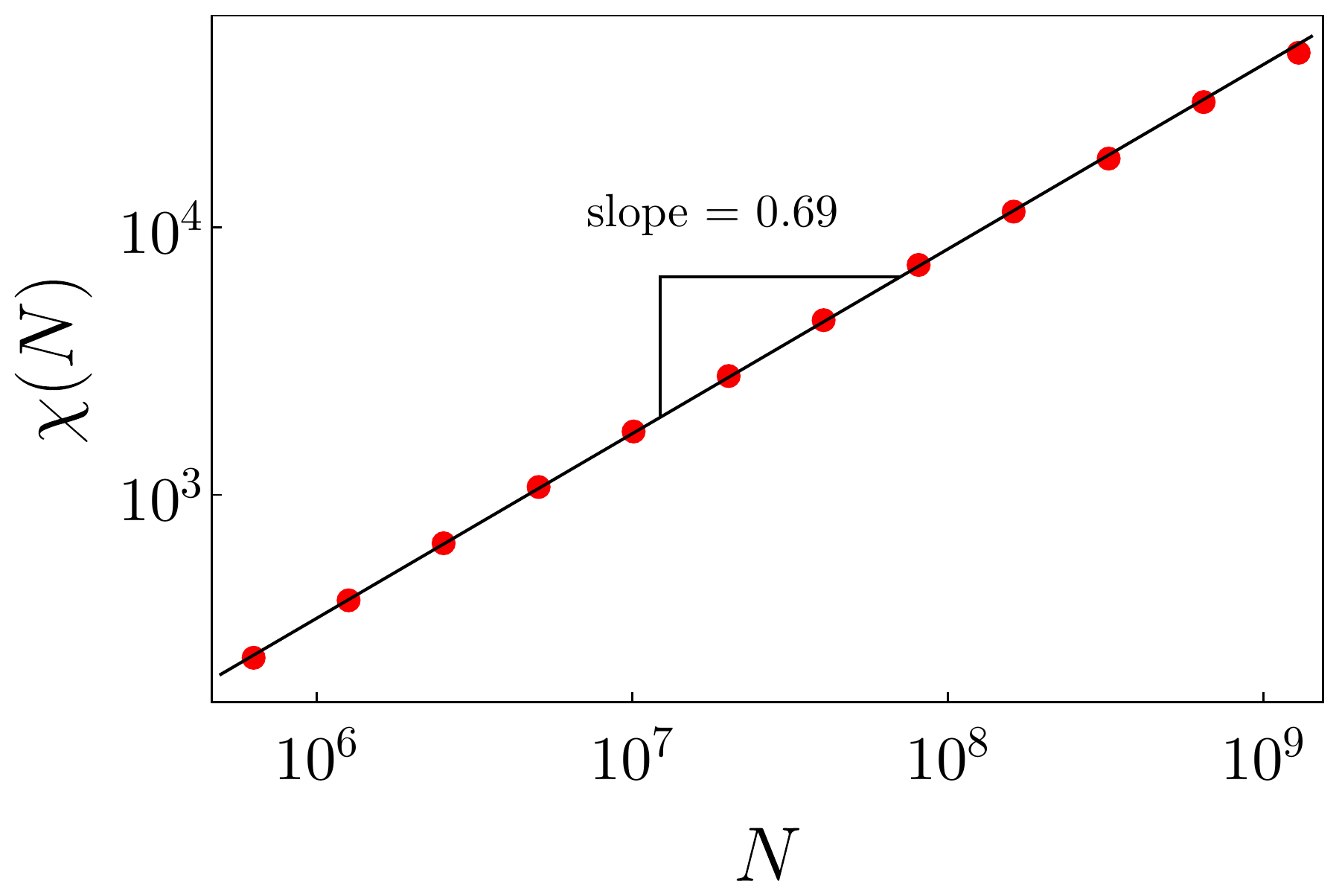}
\caption{Plot of the susceptibility $\chi$, the fluctuation of the order parameter $m_d$, as a function of the system size $ N $ at $ \kappa_c \approx 0.108021 $. $\gamma / \nubar$ is estimated to be $ \approx 0.69 $. Data are obtained for ER networks with $z=8 $. $\ro$ is taken as $\rho_c\approx 0.00747762$. }
\label{fig12}
\end{figure}

\begin{figure}[H]
\centering
\includegraphics[width=0.9\linewidth]{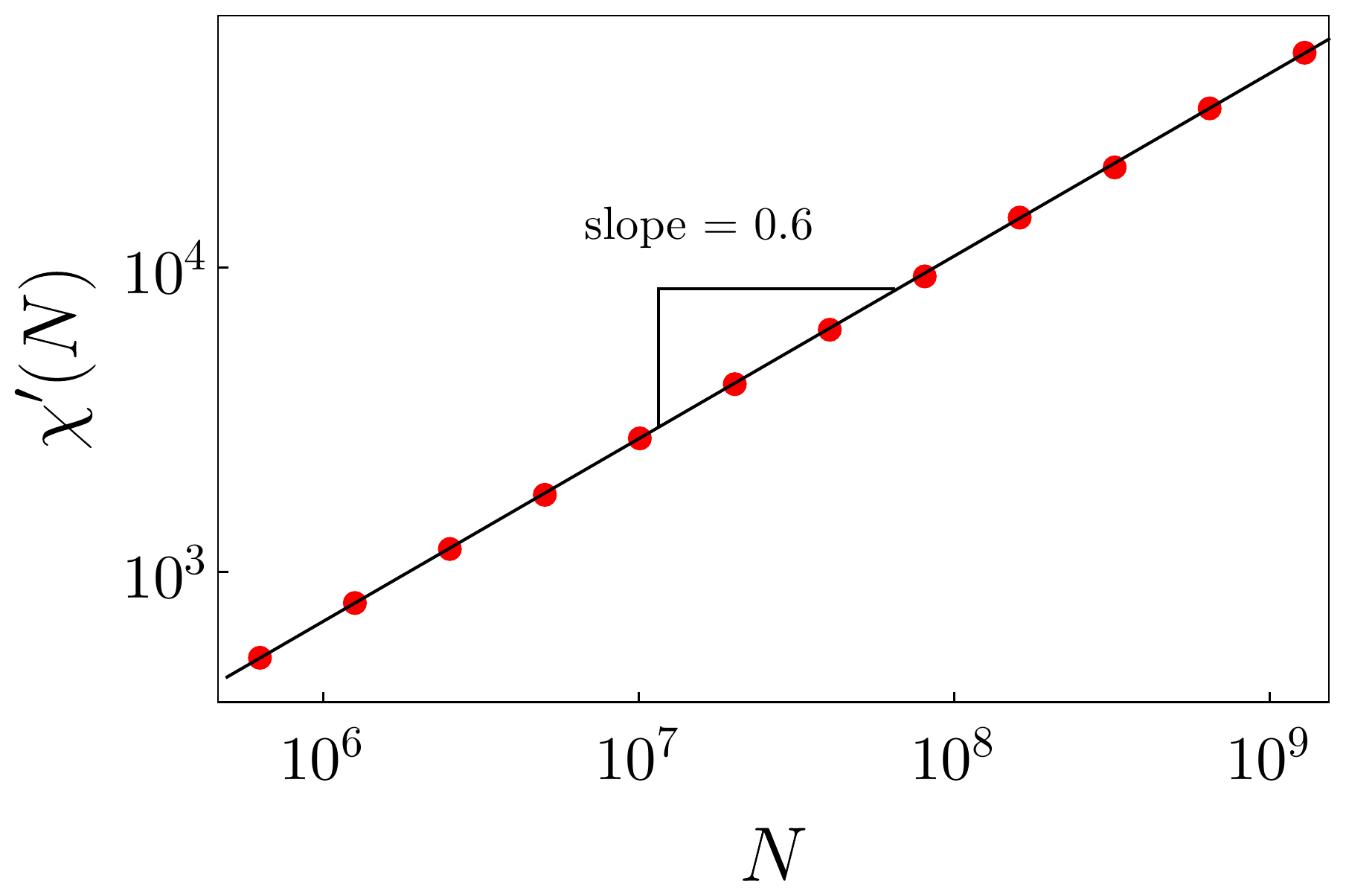}
\caption{Plot of the susceptibility $\chi^{\prime}$, the fluctuation of the order parameter $m_u$, as a function of the system size $ N $ at $ \kappa_c \approx 0.108021 $. $\gamma^{\prime} /\nubar^{\prime}$ is estimated to be $\approx 0.6 $. Data are obtained for ER networks with mean degree $z=8 $. $\ro=\rho_c$ is taken as $\approx 0.00747762$.}
\label{fig13}
\end{figure}

For $\kappa<\kappa_c$, the fluctuation of the order parameter $ \chi \equiv N( \langle m_d^2 \rangle - \langle m_d \rangle^2)$ behaves as $\sim N^{\gamma/\bar{\nu}}G[(\kappa_c-\kappa)N^{1/\bar{\nu}}] $ with a certain scaling function $G$. On the other hand, for $\kappa>\kappa_c$, we obtain that $ \chi^{\prime}\equiv N( \langle m_u^2 \rangle - \langle m_u \rangle^2)$ behaves as  $\sim N^{\gamma^{\prime}/\bar{\nu}^{\prime}}G^{\prime}[(\kappa-\kappa_c)N^{1/\bar{\nu}^{\prime}}] $ with a certain scaling function $G^{\prime}$.
We numerically obtain $ \gamma / \nubar \approx 0.69 $ (Fig.~\ref{fig12}) and $ \gamma^{\prime} / \nubar^{\prime} \approx 0.6 $ (Fig.~\ref{fig13}).

\begin{figure}[h]
\centering
\includegraphics[width=0.9\linewidth]{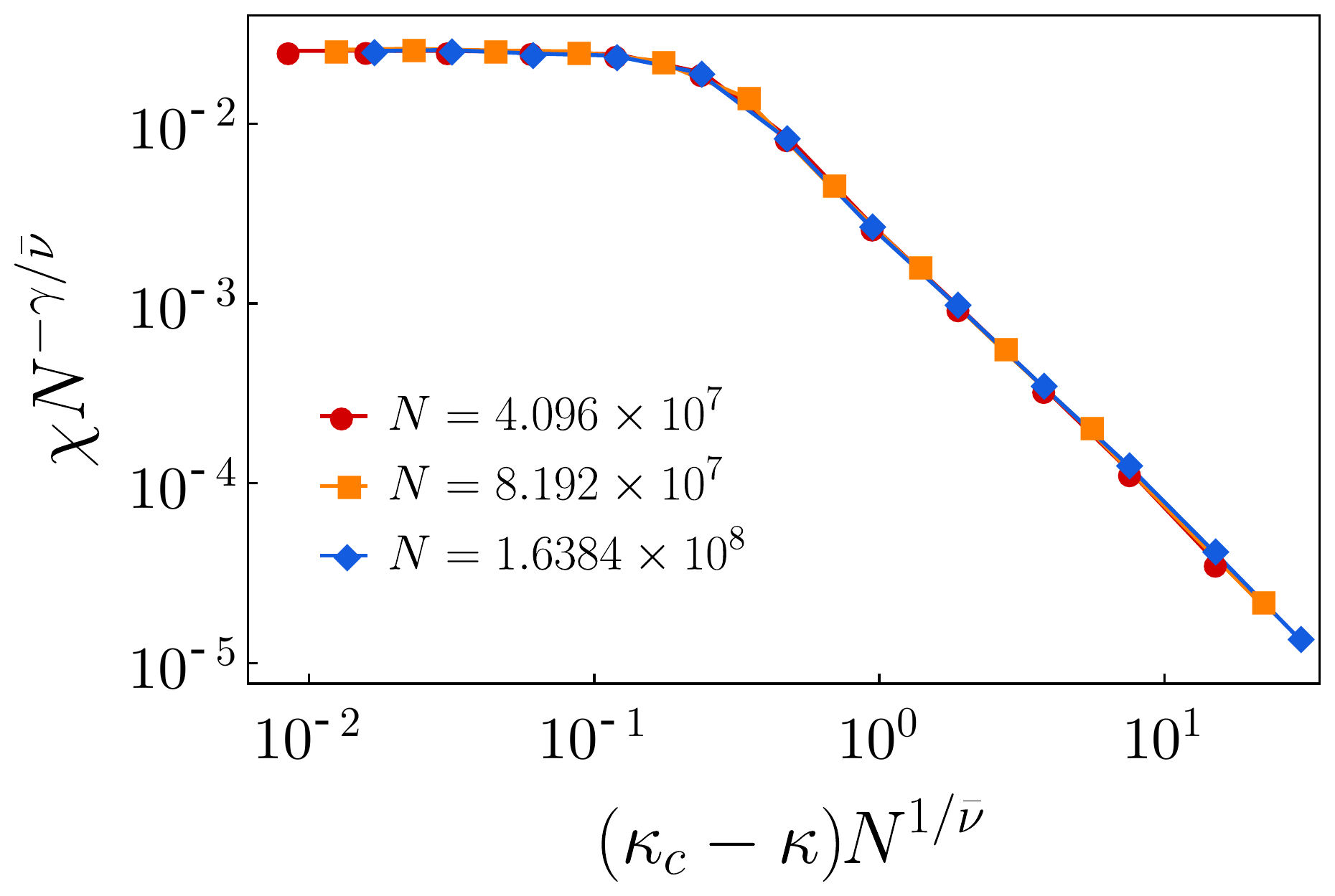}
\caption{Scaling plot of the susceptibility $\chi$ for $\kappa < \kappa_c$ in the form $\chi N^{-\gamma/\nubar}$ versus $(\kappa-\kappa_c)N^{1/\nubar}$, where $\nubar\approx 2.13$ and $\gamma\approx 1.47$ are used. Data are obtained for ER networks with mean degree $z=8 $. $\ro=\rho_c$ is taken as $\approx 0.00747762$. }
\label{fig14}
\end{figure}

\begin{figure}[h]
\centering
\includegraphics[width=0.9\linewidth]{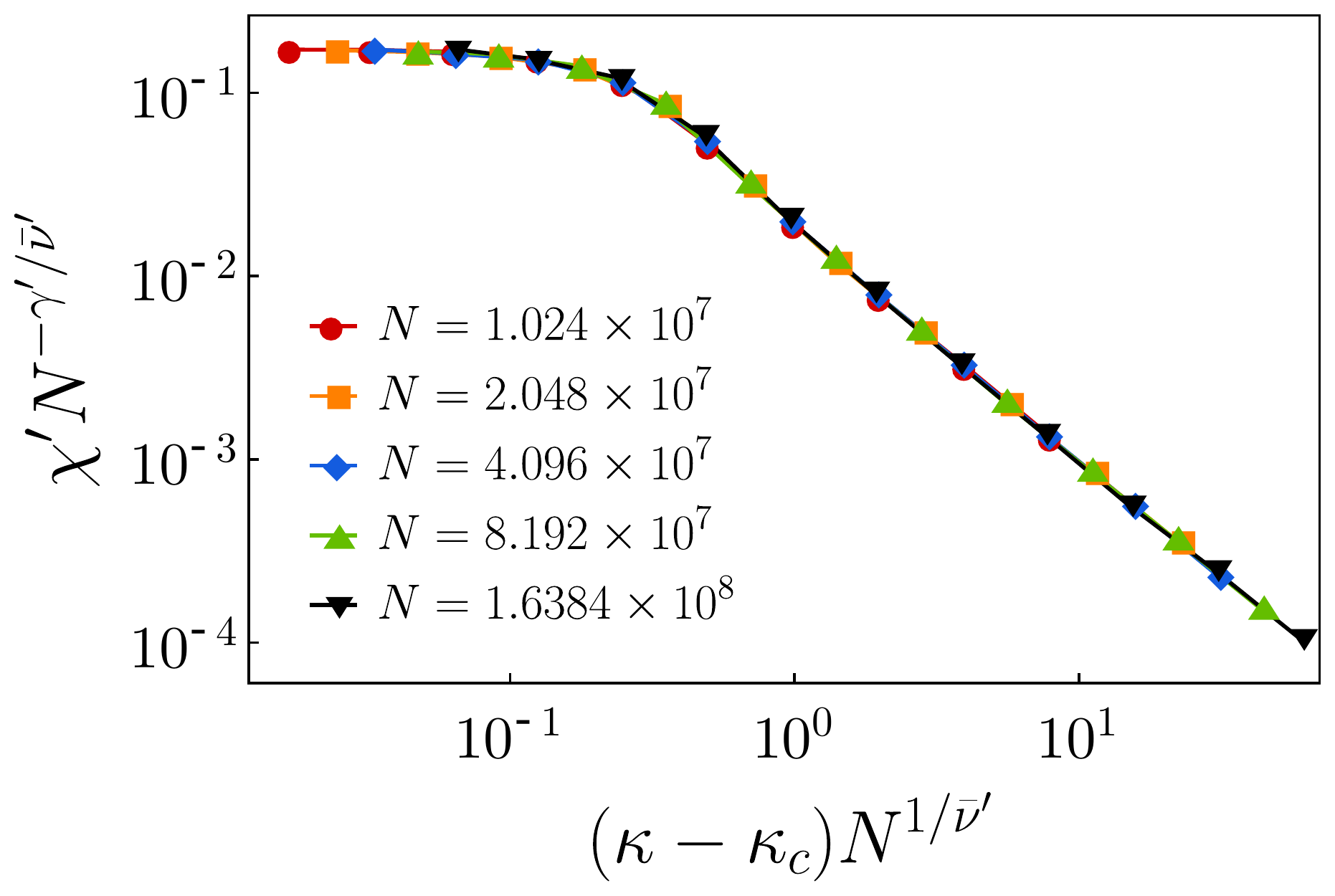}
\caption{Scaling plot of the susceptibility $\chi^{\prime}$ for $\kappa > \kappa_c$ in the form $\chi^{\prime} N^{-\gamma^{\prime}/\nubar^{\prime}}$ versus $(\kappa-\kappa_c)N^{1/\nubar^{\prime}}$, where $\nubar^{\prime}\approx 2.13$ and $\gamma^{\prime}\approx 1.28$ are used. Data are obtained for ER networks with mean degree $z=8 $. $\ro=\rho_c$ is taken as $\approx 0.00747762$. }
\label{fig15}
\end{figure}

On the basis of the numerically obtained values $\beta/\nubar\approx 0.53$ and $\gamma/\nubar\approx 0.69$, and the theoretical value $\beta=1/3$, we estimate $\nubar \approx 2.179$ and $\gamma \approx 1.5$. Those values are confirmed in Fig.~\ref{fig14} for $\chi N^{-\gamma/\bar{\nu}}$ versus $(\kappa_c -\kappa)N^{1/\bar{\nu}}$, in which the data collapse well with the choices of $\bar{\nu} \approx 2.13\pm 0.1$ and $\gamma/\bar{\nu}\approx 0.69$. The measured values of the exponents satisfy the hyperscaling relation $ (2\beta+\gamma)/\bar{\nu}\approx 0.996 $ well. Similarly, for $\kappa > \kappa_c$, on the basis of the numerical values $\beta^{\prime}/\nubar^{\prime}\approx 0.164 $ and $ \gamma^{\prime} / \nubar^{\prime} \approx 0.6 $, and the theoretical value $\beta^{\prime}=1/3$, we obtain $\nubar^{\prime}\approx 2.03 $ and $ \gamma^{\prime} \approx 1.218 $. Data for $\chi^{\prime}$ for different system sizes collapse well into a single curve with the choices of $\bar{\nu}^{\prime}=2.13 \pm 0.1$ and $\gamma^{\prime}/\bar{\nu}^{\prime}=0.6$ (Fig.~\ref{fig15}). These values yield 
$ (2\beta^{\prime}+\gamma^{\prime})/\nubar^{\prime}\approx 0.91-0.93$, which deviates slightly from the expected value of unity that would satisfy the hyperscaling relation. To obtain those results, we used the numerical values $\rho_c \approx 0.00747762$ and $\kappa_c \approx 0.108021$. We remark that $ \beta=\beta^{\prime}=1/3 $ is obtained analytically.

\section{Summary and discussion}
We investigated the properties of phase transitions in the SWIR model with a finite density $ \ro $ of initially infected seeds~\cite{janssen}. A node in the state $S$ can change its state to weakened ($W$) or infected ($I$) when it comes in contact with an infected node from the same or a different root. A weakened node can also change its state to infected ($I$) when it contacts an infected node from the same or a different root. The reaction probabilities $\kappa$ and $\mu$ in Eqs.~(\ref{si_ii}) and (\ref{si_wi}), respectively,  serve as control parameters. For convenience, we take $\kappa=\mu$. We found that for a given network, there exists a critical density of seeds $\rho_c$ such that for $\ro < \rho_c$, the order parameter, the density of nodes in state $R$ in the absorbing state, increases continuously with the critical exponent $\beta=1/2$ as $\kappa$ is increased up to a transition point $\kappa_c^-$ and then jumps to a finite value, followed by a continuous increase. Accordingly, the order parameter behaves as $m(\kappa)=m_0^{-}-b (\kappa_c^--\kappa)^{1/2}$ for $\kappa < \kappa_c^-$, where $b$ is a positive constant. At $\kappa_c^-$, the order parameter is discontinuous by $\Delta m=m_u(\kappa_c^-)-m_0^-$. Thus, the order parameter itself exhibits a hybrid phase transition. This pattern is different from that for the single-seed case, in which the order parameter jumps from $m=0$ to a finite value, and thus $\beta=0$. The fluctuation of the order parameter diverges at the transition point $\kappa_c^-$ according to a power-law with the exponent $\gamma$. For the correlation size exponent $\nubar$ measured in finite systems, we find that the hyperscaling relation $2\beta+\gamma=\nubar$ holds reasonably well.  

As $\ro$ is increased, the jump shrinks and becomes zero at $\rho_c$. For $ \ro =\rho_c $, the transition becomes continuous. We determined a complete set of critical exponents describing the phase transition at $\kappa_c$. The critical exponents are listed in Table I.

\begin{table*}
	\caption{List of the critical exponents of the SWIR models with a single seed and with multiple seeds.}	 
	\begin{center}
		\setlength{\tabcolsep}{12pt}
		{\renewcommand{\arraystretch}{1.5}
			\begin{tabular}{lcccccccc}
				\hline
				\hline
				
				& type  & $\beta$ & $\beta'$ & $\gamma$ & $\gamma'$ & $\bar{\nu}$ & $\bar{\nu}^{\prime}$  \\
				\hline
				
				$ \ro=1/N $ & single seed  & 0 & - & - & - & 3 & -\\
				
				$ 0<\ro<\rho_c $
				& multiple seeds  & 1/2 & - & $1$ & - & $2$ & - \\

				$\ro = \rho_c$
				& multiple seeds & 1/3 & 1/3 & $1.47\pm0.05$ & $1.28\pm0.05$ & $2.13\pm0.1$ & $2.13\pm0.1$ \\		
				
				\hline
				\hline
		\end{tabular}}
	\end{center}
\end{table*}

\hskip 0.5cm
\begin{acknowledgments}
This work was supported by the National Research Foundation of Korea by Grant No. NRF-2014R1A3A2069005.
\end{acknowledgments} 

\appendix

\section{Derivation of the critical exponent $\beta$ at $\rho_c$} 
Here we introduce an analytical method to determine the critical exponents $\beta=1/3$ at $\ro=\rho_c$. It is already noted in Sec. III-B that for $\ro=\rho_c$,
\begin{equation}\label{eq:saddle}
F(\kappa_c,m_0)=\dfrac{dF}{dm}\bigg|_{\kappa_c,m_0}
=\dfrac{d^2F}{dm^2}\bigg|_{\kappa_c,m_0}=0.
\end{equation}
We consider a line of the solution $F(\kappa,m)=0$ near $(\kappa_c,m_0)$ by expanding $F(\kappa_c+\delta\kappa,m_0+\delta m)$ as
\begin{widetext}
	\begin{equation}\label{eq:expansion}
	F(\kappa_c+\delta\kappa, m_0+\delta m) \simeq 
	\dfrac{\partial F}{\partial \kappa}\bigg|_{\kappa_c,m_0}(\delta \kappa) 
	+\dfrac{1}{6}\dfrac{\partial^3 F}{\partial m^3}\bigg|_{\kappa_c,m_0}(\delta m)^3+\dots=0 
	\end{equation}
\end{widetext}
where only nonzero terms are considered. Since $\delta \kappa$ and $(\delta m)^3$ are two lowest terms in Eq.~(\ref{eq:expansion}) and their coefficients have the opposite sign to each other, $\delta m \sim (\delta \kappa)^{1/3}$ when $ \delta \kappa \ll 1 $. Thus for both cases of $ \kappa<(>)\kappa_c $, the critical exponents $ \beta=\beta^{\prime}=1/3 $.

\end{document}